\input{aipcheck}

\documentclass[
    ,final            % use final for the camera ready runs
  ]
  {aipproc}

\layoutstyle{6x9}

\begin{document}

\title{A Planetary Companion around a Metal-Poor Star with Extragalactic Origin}


\author{Johny Setiawan}{
  address={Max-Planck-Institut f\"ur Astronomie, K\"onigstuhl 17, 
  69117~Heidelberg, Germany}
}

\author{Rainer Klement}{
  address={Max-Planck-Institut f\"ur Astronomie, K\"onigstuhl 17,
  69117~Heidelberg, Germany}
}

\author{Thomas Henning}{
  address={Max-Planck-Institut f\"ur Astronomie, K\"onigstuhl 17,
  69117~Heidelberg, Germany}
   % additional visiting address
}
\author{Hans-Walter Rix}{
  address={Max-Planck-Institut f\"ur Astronomie, K\"onigstuhl 17,
  69117~Heidelberg, Germany}
   % additional visiting address
}
\author{Boyke Rochau}{
  address={Max-Planck-Institut f\"ur Astronomie, K\"onigstuhl 17,
   69117~Heidelberg, Germany}
   % additional visiting address
}
\author{Tim Schulze-Hartung}{
  address={Max-Planck-Institut f\"ur Astronomie, K\"onigstuhl 17,
  69117~Heidelberg, Germany}
   % additional visiting address
}
\author{Jens Rodmann}{
  address={European Space Agency, Space Environment and Effects Section, ESTEC, Keplerlaan 1, 2201 AZ Noordwijk, The Netherlands}
   % additional visiting address
}

\begin{abstract}
We report the detection of a planetary companion around \mbox{HIP 13044}, 
a metal-poor star on the red Horizontal Branch. The detection
is based on radial velocity observations with FEROS, 
a high-resolution spectrograph at the 2.2-m MPG/ESO telescope, located at ESO 
La Silla observatory in Chile. The periodic radial velocity 
variation of $P=16.2$ days can be distinguished
from the periods of the stellar activity indicators.
We computed a minimum planetary 
mass of 1.25 M$_\mathrm{Jup}$ and an orbital semi-major axis 
of 0.116 AU for the planet.
This discovery is unique in three aspects: 
First, it is the first planet detection around a star with a metallicity 
much lower than few percent of the solar value; second, 
the planet host star resides in a stellar evolutionary stage 
that is still unexplored in the exoplanet surveys; 
third, the star \mbox{HIP 13044} belongs to one of the most significant stellar 
halo streams in the solar neighborhood, 
implying an extragalactic origin of the planetary system \mbox{HIP 13044} in 
a disrupted former satellite of the Milky Way. 
\end{abstract}

\maketitle

%%%%%%%%%%%%%%%%%%%%%%%%%%%%%%%%%%%%%%%%%%%%
%% MAINMATTER
%%%%%%%%%%%%%%%%%%%%%%%%%%%%%%%%%%%%%%%%%%%%

\section{Introduction}
Several hundred extrasolar planets beyond our Solar system 
have been discovered in the last two decades. 
So far, the majority are companions around solar-like stars. 
Compared to these discoveries, there is only a small number of planet 
detections in the late evolutionary state, like around Red 
Giant Branch (RGB) stars and pulsars. Even more so, the phase directly after the 
RGB stage, the Horizontal Branch (HB), is still unexplored. 

Besides the evolutionary stage, a star's chemical composition seems to 
be a major indicator of its probability for hosting a planet. Early  
studies, e.g., \cite{gon97}, showed that giant planet host stars are metal-rich. 
This finding is supported by the large exoplanet search surveys around main-sequence stars 
that reported the planet-metallicity connection \citep{san04,val05}. 
Moreover, the survey of metal-poor main-sequence stars e.g.\ \cite{soz09} 
has found no evidence for Jovian planets. 

Until now, only very few planets have been discovered around stars 
with metallicities as low as [Fe/H]=\, $-1$, i.e.\ 10\% of the Sun's metallicity. 
The discovery of PSR B1620 b, a Jovian planet orbiting a pulsar 
in the core of the metal-poor globular cluster M4 ([Fe/H]=$-1.2$), 
implies, however, that planets might have formed around 
metal-poor stars \citep{ford00,sig03}.

\section{The star \mbox{HIP 13044}}
\mbox{HIP 13044} is a star of spectral type F2 (SIMBAD). 
It resides in the red part of the Horizontal Branch (RHB), 
which is separated from the blue and extreme Horizontal Branch (BHB and EHB) 
stars by the RR Lyrae instability strip (Fig.~1). 
Table~1 summarizes the fundamental stellar parameters.

\begin{figure}[t]
  \includegraphics[width=0.8\textwidth]{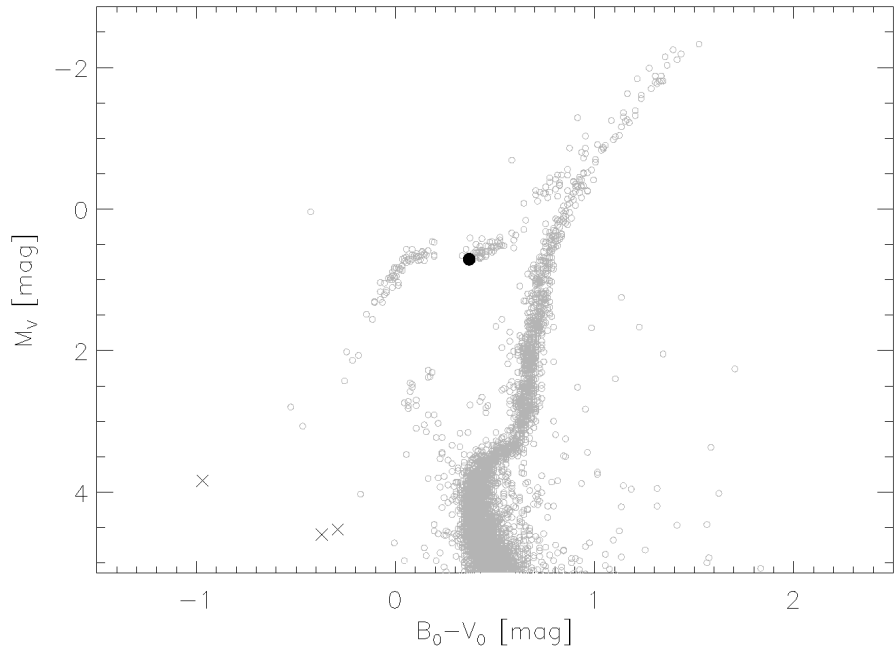}
 \caption{Location of \mbox{HIP 13044} in a M$_V$ vs. $B-V$ color-magnitude 
 diagram (CMD) shown as a black dot superimposed to the CMD of Messier 3 (grey open circles) 
 based on the photometry by \cite{buo94}. Further candidates for post RGB stars hosting 
 planets/brown dwarf, V391 Peg, HW Vir and HD 149382 \cite{sil07,lee09,gei09} are 
 displayed as crosses.}
\label{fig:1}
\end{figure}

\begin{table}[t]
\caption{Stellar parameters of \mbox{HIP 13044}}

\vspace{0.5cm}
\begin{tabular}{llll}
\hline  
  Parameter      &  Value		&  Unit   	& Reference\tablenote{[1]:\cite{bee90};[2]:\cite{car86};[3]:\cite{car08a};
[4]:\cite{roe10};[5]:\cite{chi00};[6]:\cite{car08b}}
 \\
 \hline
 Spectral type      &  F2		&     		& {\sc SIMBAD}\\
 $m_{V}$            &  9.94 		& mag 		& Hipparcos\\
 distance           &  701$\pm$20 	& pc  		& [1],[2],[3]\\
 $T_{\mathrm{eff}}$ &  6025$\pm$63	& K     	& [3],[4]\\
 $R_*$		    &  6.7$\pm$0.3 	& R$_{\odot}$ 	& [3]\\

 $\log{g}$	    &  2.69$\pm$0.3	&       	& [3]\\
 $m$ 		    &  0.8$\pm$0.1	& M$_{\odot}$ 	& this work\\
 $[Fe/H]$           &  -2.09$\pm$0.26	& [Fe/H]$_{\odot}$ & [1],[4],[5],[6]\\
 $v \sin{i}$	    &  8.8$\pm$0.8   	& $\mathrm{km\,s}^{-1}$ & [3] \\
 		    &  11.7$\pm$1.0   	& $\mathrm{km\,s}^{-1}$ & this work\\
 \hline 
 \end{tabular}

\end{table}

The stellar mass has been inferred from the knowledge of 
the stellar radius $R_\ast$ and surface gravity $\log{g}$. 
We calculated a stellar mass of $0.8\pm0.1 M_{\odot}$. 
\cite{car08a} has measured $v \sin{i}= 8.8\, \mathrm{km\,s}^{-1}$, 
whereas we obtained $v \sin{i}= 11.7\, \mathrm{km\,s}^{-1}$. 
The discrepancy between the two results is probably 
caused by the different methods used. \cite{car08a} used a 
Fourier transform method, whereas we used a cross-correlation 
technique to measure the $v \sin{i}$ \cite{set04}.
For this work we adopt $v \sin{i}= 10.25\pm2.1 \, \mathrm{km\,s}^{-1}$ 
which is the mean value of both measurements.

The star is known to be a member of the Helmi stellar stream, a group of stars that 
share similar orbital parameters that, however, stand apart from the orbits of 
the bulk of all the other stars. 
The stream stars once were bound to a satellite galaxy of the Milky Way \citep{hel99,chi00}
that was tidally disrupted 6--9 Gyr ago \citep{kep07}. 
The stream membership of \mbox{HIP 13044} has been confirmed by
several authors \citep{ref05,kep07,roe10}. 
\mbox{HIP 13044} shares the property of other stream members, like the low iron abundances 
([Fe/H]$=-1.8$ for 33 stream members) \citep{kep07,kle09} 
and a chemical similarity to typical inner halo stars \cite{roe10}. 
An extensive abundance analysis of \mbox{HIP 13044} has been also 
presented in \cite{roe10}.

\section{Observations}
Previous radial velocity (RV) measurements of \mbox{HIP 13044} showed 
a systematic velocity of about 300 $\mathrm{km\,s}^{-1}$ with respect to the Sun, 
indicating that the star belongs to the stellar halo \citep{car86}. 

We observed \mbox{HIP 13044} from September 2009 until July 2010 with FEROS
\citep{kau00},
a high-resolution spectrograph ($R$ = 48\,000) attached at the 2.2 meter
Max-Planck Gesellschaft/European Southern Observatory (MPG/ESO) telescope, 
located at the ESO La Silla observatory in Chile. 
To measure the RV values of \mbox{HIP 13044} we used a cross-correlation technique, where the
stellar spectrum is cross-correlated with a numerical template (mask) 
designed for stars of the spectral type F0.

\subsection{Radial velocity variation}
The variation of the RV between our observations has a 
semi-amplitude of 120 $\mathrm{m\,s}^{-1}$.
In order to search for periodic variations, we used periodogram analysis techniques, 
which are capable of treating missing values and unevenly spaced time points.
The Generalized Lomb Scargle (GLS) periodogram \citep{zec09} 
reveals a significant RV periodicity at $P=16.2$ days with 
a False Alarm Probability (FAP) of $5.5\times10^{-6}$ (Fig.~2). 
Additional analysis by using a Bayesian algorithm \citep{gre05} yielded 
also a period around 16 days. 

Such RV variation can be induced in principle either by an unseen orbiting companion, 
by moving/rotating surface inhomogeneities or by non-radial 
stellar pulsations. Exploring both stellar rotational modulation 
and pulsations is critical when probing the
presence of a planetary companion, since they can produce a similar or even
the same RV variation, mimicking a Keplerian motion.

\begin{figure}
  \includegraphics[width=0.8\textwidth]{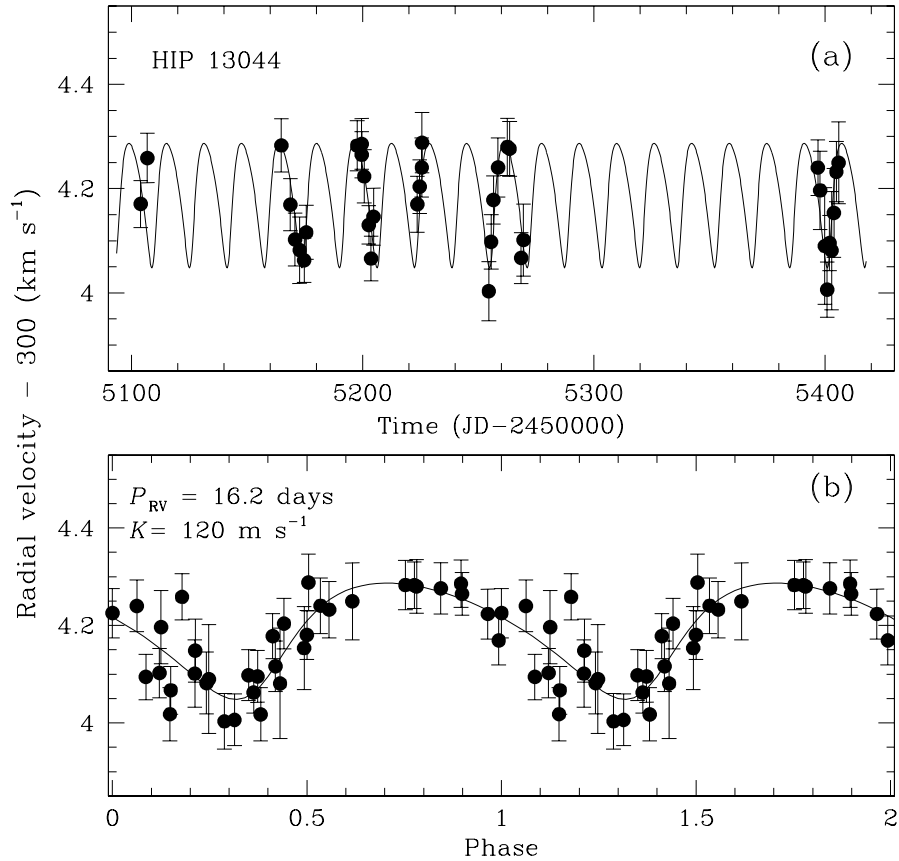}
 \caption{RV variation of \mbox{HIP 13044} is shown in the upper panel. 
 The lower panel shows the RV variation phase-folded with $P=$16.2 days. 
 The amplitude of the RV variation is $120\,\mathrm{m\,s}^{-1}$.}
\label{fig:2}
\end{figure}

\subsection{Stellar activity}
\subsubsection{Bisector and Ca II}
A well-established technique to detect stellar rotational modulation 
is to investigate the line profile asymmetry or bisector \citep{gray08} and Ca II lines. 
Surface inhomogeneities, such as star spots 
and large granulation cells, will produce asymmetry in the spectral line
profiles. Instead of measuring the bisectors, one can equivalently use 
the bisector velocity spans (BVS) to search for rotational 
modulation \citep{hat96}.
Adopting this technique, we have measured BVS from the stellar spectra. 
We found only a weak correlation between BVS and RV (correlation
coefficient ={\bf $-0.13$}). Interestingly, we found that the BVS variation shows a clear periodicity 
with $P=$5.02 days (Fig.~3). No period around 16 days is found in the BVS variation.  

\begin{figure}
\includegraphics[width=0.8\textwidth]{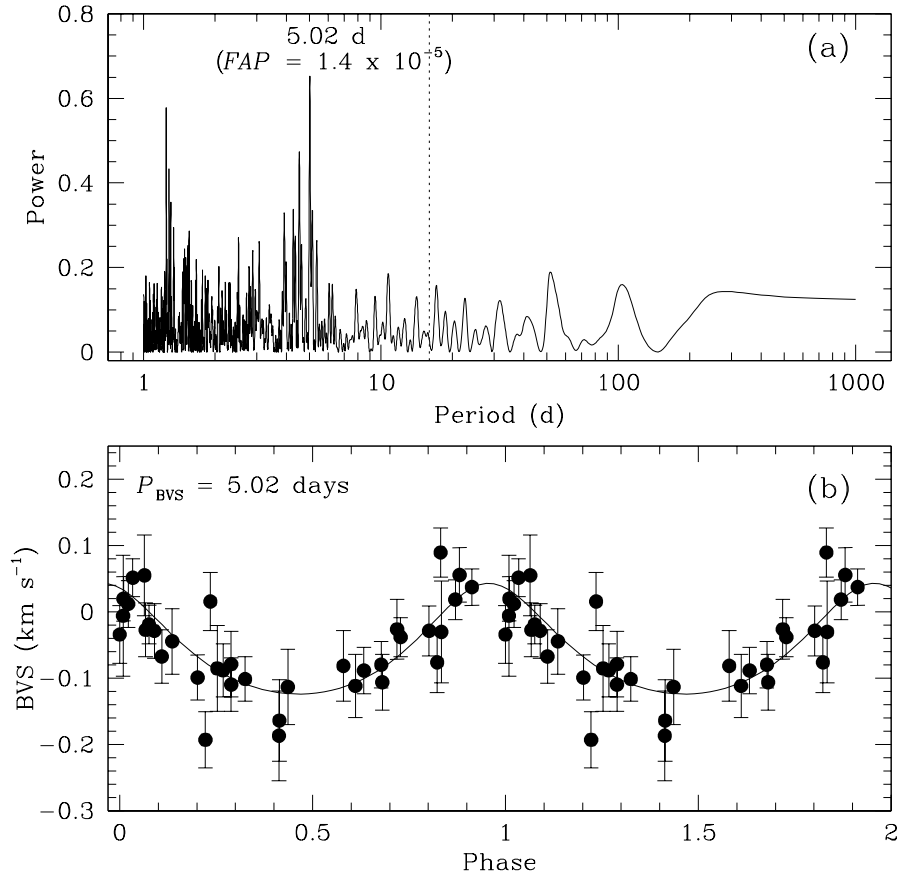}
\caption{Periodogram analysis of the BVS variation showing a periodicity 
of 5.02 days with a FAP of $1.4\times 10^{-5}$ (upper panel). 
The lower panel shows a phase-folded plot of the BVS variation. 
The semi-amplitude of the variation is $\sim70\mathrm{m\,s}^{-1}$.}
\label{fig:3}
\end{figure}

In addition to the BVS analysis, we investigated the variation of the 
Ca II $\lambda$849.8 line, which is one of the Ca II infrared triplet lines. 
From the observed Ca II 849.8 equivalent width variations we computed 
a mean period of 6.05 days, 
which is in the same order of the period of the BVS variation. 
We adopt the mean period of both stellar activity indicators, 
$P_{\mathrm{rot}}=5.53\pm0.73$ days, 
as the stellar rotation period of \mbox{HIP 13044}.

\subsubsection{Photometric observations}
We found evidence for stellar pulsations of \mbox{HIP 13044} 
from the photometric measurements with Hipparcos \citep{per97} 
and SuperWASP \citep{pol06}. 
While the Hipparcos data shows only one marginal significant periodicity 
of 7.1 hours, the SuperWASP data
shows two significant periodicities of $P_1=1.39$ days and $P_2=3.53$ days 
(FAP=$5\times 10^{-4}$ and $2\times 10^{-4}$). 
These two signals, however, are harmonic to each other ($1/P_1+1/P_2=1$).

The timescale of the Hipparcos photometric variation 
is similar to oscillation periods found in evolved stars in the 
metal-poor globular cluster NGC 6397 \citep{ste09}. 
According to the empirical relation by \cite{kje95} 
we estimated a oscillation frequency of 53 $\mu$Hz that 
corresponds to a period of 5.2 hours. 
The timescale of the derived periodicity is further consistent 
with theoretical models of pulsations of HB stars
\citep{xio98}. 

However, the periodicities found in the SuperWASP data 
do not seem to support the models by \cite{kje95} and
\cite{xio98}. It is possible, that these are high-order overtones of the 
stellar pulsations. Although the true nature of the 
stellar oscillations of \mbox{HIP 13044} 
is still not yet fully understood, based on the results of the 
SuperWASP photometric observations, stellar pulsations 
can be ruled out as the source of the 16.2 days periodic RV variation.

\section{Planetary companion}
Based on the above results the best and only viable explanation for the 
$\sim$16 days period is the presence of an unseen companion. 
We computed the orbital solution of the companion, 
as given in Table~2.

\begin{table}[!t]
 \caption{Orbital parameters of \mbox{HIP 13044} b}
   \begin{tabular}{l@{\qquad}ll}
 \hline 
 $P$      	    &  16.2    $\pm$ 0.3 	    & days  \\
 $K_{1}$            &  119.9   $\pm$ 9.8	    & $\mathrm{m\,s}^{-1}$\\
 $e$                &  0.25    $\pm$ 0.05	    &	    \\
 $\omega$           &  219.8   $\pm$ 1.8	    & deg   \\
 $JD_{0}-2450000$   &  5109.78 $\pm$ 0.02	    & days  \\
 $\chi^{2}$ 	    &  32.35 		    	    & $\mathrm{m\,s}^{-1}$\\
 $rms$         	    &  50.86     		    & $\mathrm{m\,s}^{-1}$\\
 $m_{2} sin{i}$     &  1.25    $\pm$ 0.05	   & M$_{\mathrm{Jup}}$\\
 $a$	            &  0.116   $\pm$ 0.01	   & AU    \\
 \hline
 
 \end{tabular}
\end{table}

The minimum mass of the companion lies securely in the planetary mass domain, 
even with a plausible $\sin{i}$ uncertainty. 
With an eccentricity of 0.25 and a semi-major axis of 0.116 AU, the planet
has a periastron distance of about 0.087 AU which is $\approx$2.8 times of 
the present stellar radius. 
The non-circular orbit ($e=$0.25) is, however, not expected for 
a close-in giant planet around a post RGB star. It is possible that 
there is a third body in the system that affects the planetary orbit. 

In the case of \mbox{HIP 13044}, the original orbit could have been disturbed 
or changed during the evolution of the star-planet-system, 
in particular during the RGB phase \citep{sok98}. 
Interestingly, the orbital period of \mbox{HIP 13044} b is close to three times
the stellar rotation period. Such planetary systems are particularly interesting to study 
the star-planet interaction e.g., \cite{shk08}.

\section{Discussion}
The discovery of a planetary companion around \mbox{HIP 13044} is a 
particularly interesting finding since the star has a very low metallicity 
([Fe/H]$\approx-2.1$).
This is the lowest stellar metallicity among all known planet host stars
so far (Fig.~4).

\begin{figure}
  \includegraphics[width=0.76\textwidth]{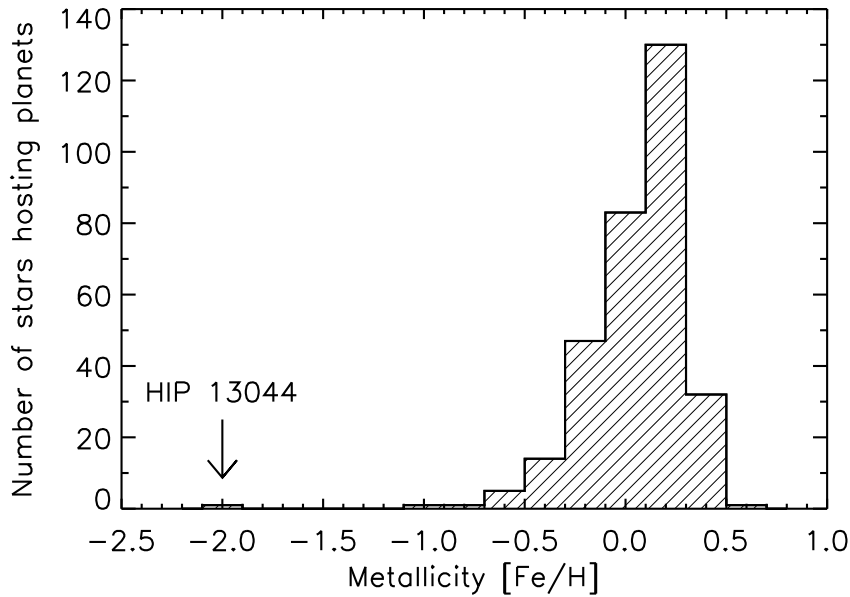}
 \caption{Distribution of the metallicity [Fe/H] of planet host stars. \mbox{HIP 13044}, 
 with a mean metallicity estimate of [Fe/H]=-2 is far more metal-poor 
 than any previously known exo-planet host star.}
\label{fig:4}
\end{figure}

It might be the case that initially, the star had a higher metallicity in the
planet formation phase and subsequently lost the metallicity during its
evolution, especially in the giant phase via complex mechanisms. 
Such a mechanism could be the incorporation of heavy elements 
into dust grains followed by their separation from the atmosphere \citep{mat92}.
However, given the star's membership of the Helmi stream, in which
the most metal-rich subdwarfs known so far have [Fe/H]$\sim-1.5$ 
\citep{kle09}, we do not expect its initial Fe abundance to exceed this value. 

So far, there are only very few planet/brown dwarf 
detections around post RGB stars beyond the pulsar planets, 
e.g.\ \mbox{V391 Peg} \citep{sil07}, \mbox{HW Vir} \citep{lee09} 
and \mbox{HD 149382} (\citealp{gei09}, Fig.~1); 
%see also \citet{sil2010} for a complete list.
see also Silvotti et al.\ (these proceedings) for a complete list.
These are, however, substellar companions around 
subdwarf-B or Extreme Horizontal Branch (EHB) stars, i.e, 
the nature of the host stars is different from RHB stars like \mbox{HIP 13044}. 
Contrary to planet search surveys around RGB stars, such as G and K giants, 
that have been carried out extensively \citep{hat93,set04,doe07}, 
HB stars are still unexplored. 

While at least 150 main-sequence stars are known to bear close-in ($a=0.1$
AU) giant planets, so far no such planets have been reported
around RGB stars. A possible explanation is that the inner planets have
been engulfed by the star when the stellar atmosphere expands during the
giant phase. Theoretical models \citep{liv83,sok98,bea10} predict that massive giant  
planets can survive in the RGB phase under certain circumstances. 
In particular, the outer planets can rapidly spiral in into closer orbits, 
if the star has reached its Roche lobe during the giant phase.
However, for an adiabatic stellar mass loss, the planetary orbits 
will move out. Moreover, velocity kicks can perhaps increase eccentricity if the mass-loss 
is asymmetric \citep{hey07}.

Finally, the origin of the planetary system \mbox{HIP 13044} is also of high
interest. As a member of the Helmi stream, \mbox{HIP 13044} most probably 
has an extragalactic origin. This implies that the host star would have had 
a quite different history than the majority of known planet host stars. 
\mbox{HIP 13044} was probably
attracted to the Milky Way several Gyr ago. Before that, it
could have belonged to a satellite galaxy of the Milky Way similar to
Fornax or the Sagittarius dwarf spheroidal galaxy \citep{hel99}. 
The fascinating consequence is that \mbox{HIP 13044} b seems to be 
the first detected planet with a non-Galactic origin.

%\begin{theacknowledgments}
%We also thank W. Wang, C. Brasseur, R. Lachaume, 
%M. Zechmeister \& D. Fedele for the spectroscopic observations
%with FEROS. We would like to thank Dr. M. Perryman, Dr. Ea. Bear, Dr.
%N. Soker and Dr. P. Maxted for the fruitful discussion, comments and suggestions 
%to improve this paper. 
%\end{theacknowledgments}



\bibliographystyle{aipproc}   % if natbib is available

%%% \bibliography{setiawanbib}

 %{\typeout{}
 % \typeout{******************************************}
 % \typeout{** Please run "bibtex \jobname" to optain}
 % \typeout{** the bibliography and then re-run LaTeX}
 % \typeout{** twice to fix the references!}
 % \typeout{******************************************}
 % \typeout{}
 %}

%%%%%%%%%%%%%%%%%%%%%%%%%%%%%%%%%%%%%%%%%%%
%% The following lines show an example how to produce a bibliography
%% without the help of the BibTeX program. This could be used instead
%% of the above.
%%%%%%%%%%%%%%%%%%%%%%%%%%%%%%%%%%%%%%%%%%%

\end{document}